\documentclass{eas}
\usepackage{graphicx}

\usepackage{natbib}
\usepackage{amsmath}

\begin{document}

%%-----------------------------
%%      the top matter
%%-----------------------------
\title{Modelling the chemical evolution of star forming filaments}
\runningtitle{Chemistry in star forming filaments}
\author{D. Seifried}\address{1. Physikalisches Institut, Universit\"at zu K\"oln, Z\"ulpicher Str. 77, 50937 K\"oln, Germany}
\author{S. Walch}\sameaddress{1}
\begin{abstract}
We present simulations of star forming filaments incorporating -- to our \mbox{knowledge --} the largest chemical network used to date on-the-fly in a 3D-MHD simulation. The network contains 37 chemical species and about 300 selected reaction rates. For this we use the newly developed package KROME \citep{Grassi14}. Our results demonstrate the feasibility of using such a complex chemical network in 3D-MHD simulations on modern supercomputers. We perform simulations with different strengths of the interstellar radiation field and the cosmic ray ionisation rate and find chemical and physical results in accordance with observations and other recent numerical work.
\end{abstract}

\maketitle

\section{Introduction}

Modelling the chemical evolution of the gas during the process of star formation on various scales is a numerically and theoretically challenging task. Several authors have incorporated reduced chemical networks in their 3D, magneto-hydrodynamical (MHD) simulations \citep[e.g.][]{Clark13,Walch15}. The newly designed chemistry package KROME \citep{Grassi14} is a versatile package which allows the user to incorporate chemistry in MHD simulations in a very efficient manner and simultaneously guarantees the freedom to choose \textit{any} desired network as well as its associated cooling and heating processes. Furthermore, recently the importance of filamentary structures in star forming clouds has been re-emphasized \citep[e.g.][]{Andre10}, which make them an optimal target to test the feasibility of using a detailed chemical network in fully self-consistent 3D-MHD simulations.

\section{Initial conditions, FLASH solver, and chemical network}
\label{sec:IC}

The simulations presented here use the same initial conditions as those presented in \citet{Seifried15}. The simulated filaments have an initial width of about 0.1 pc, a length of 1.6 pc, and a mass per unit length of 75 M$_{\odot}$/pc. The initial magnetic field strength is 40 $\mu$G, in agreement with recent observations \citep[e.g.][]{Sugitani11}. The field is oriented either perpendicular or parallel to the filament. We perform several simulations with varying strengths of the interstellar radiation field (ISRF) and the cosmic ray ionisation rate (CRIR).

In order to model the chemical evolution of the gas, we use the KROME package \citep{Grassi14}. The network used contains 37 species and 287 reactions and is designed to model the formation of CO and H$_2$ in great detail (also including the formation of H$_2$ on dust). We calculate the attenuation of the ISRF as well as the self-shielding factors for H$_2$ and CO formation in the FLASH code with the TreeCol algorithm \citep{Clark12,Wunsch15}. We take into account all relevant cooling and heating mechanism present in the interstellar medium, which also includes the cooling via line emission of CO and is isotopologues $^{13}$CO and C$^{18}$O as well as a detailed calculation of the dust temperature, $T_\text{dust}$. A more detailed description of the chemistry network can be found in \citet{Seifried15b}.

\section{Results}
\label{sec:results}

\subsection{Time evolution of a fiducial run}

\begin{figure}
 \includegraphics[width=0.6\linewidth]{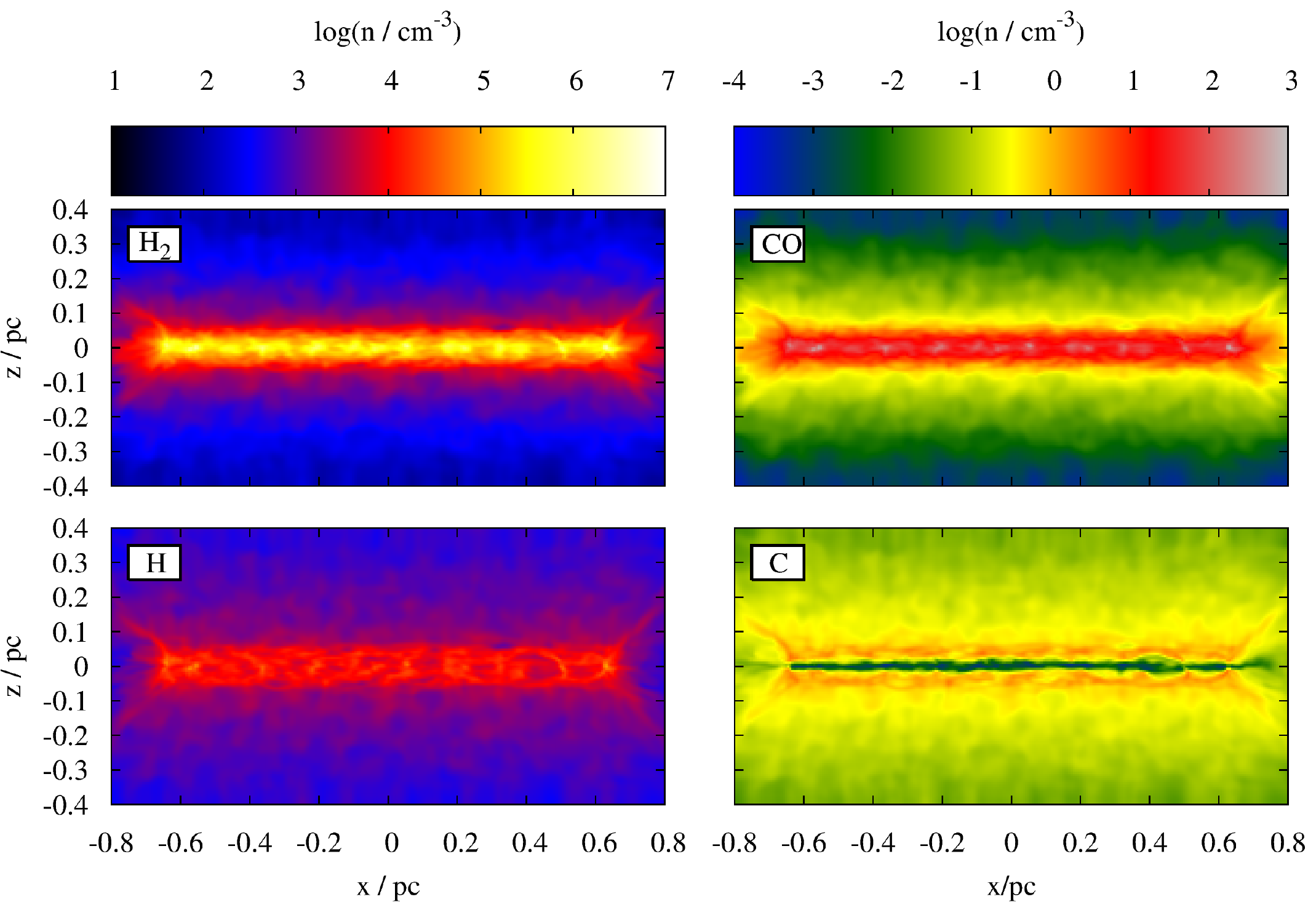} \vspace{0.5cm}
 \includegraphics[width=0.6\linewidth]{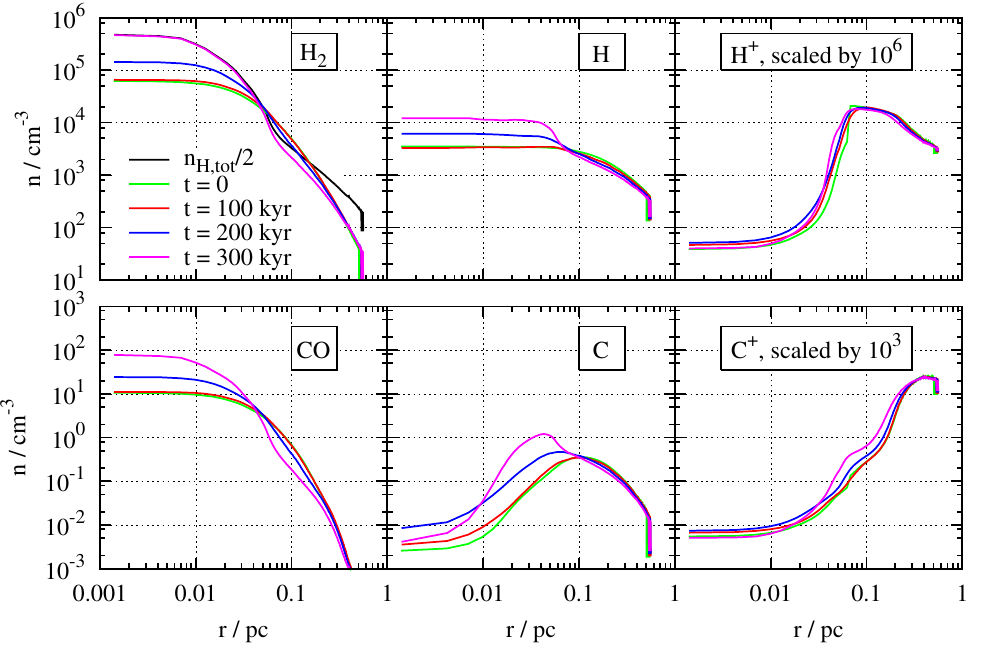}
 \caption{Left: Spatial distribution of  H, H$_2$, C, and CO at the end of the simulation with a parallel magnetic field, G$_0$ = 1.7, and a CRIR of 1.3 $\times$ 10$^{-17}$ s$^{-1}$ along a slice through the centre of the filament. Right: From left to right: Radial dependence of the density of  H$_2$, H, H$^+$, and CO, C, and C$^+$ at \mbox{t = 0, 100, 200, and 300 kyr} after the start of this simulation. The black line in the upper left panel shows the total hydrogen density divided by two.}
 \label{fig:species}
\end{figure}

In the following, we present the time evolution of the run with a parallel magnetic field, G$_0$ = 1.7, and a CRIR of 1.3 $\times$ 10$^{-17}$ s$^{-1}$. In the left panel of Fig.~\ref{fig:species} we plot the spatial distribution of H, H$_2$, C, and CO at the end of the simulation, which reveals some differences in the radial distribution. Whereas H$_2$ and CO are concentrated towards the centre of the filament, H and C are more extended. 

The radial density profiles of H$_2$, H, H$^+$, CO, C, and C$^+$ at 4 different times are shown in the right panel of Fig.~\ref{fig:species}. The constant increase of the central number densities of H, H$_2$, C, and CO over time is caused by the contraction of the filament along the radial direction. On the other hand, $n_\text{H$^+$}$ and $n_\text{C$^+$}$ remain almost unchanged over time as they recombine when the filament becomes denser. In the upper left panel we also show the number density of all H atoms (divided by two in order to be comparable to $n_\text{H$_2$}$). In the centre of the filament most of the hydrogen is bound in H$_2$, only outside $\sim$ 0.1 pc, where the black and purple line start to differ, hydrogen mainly occurs in atomic or ionised form. Hence, there is a gradual conversion of H$^+$ over H to H$_2$ towards the centre of the filament as well as of C$^+$ over C to CO, with $n_\text{C}$ exceeding $n_\text{CO}$ at $\sim$ 0.1 pc as well.

\subsection{Impact of the ISRF and the CRIR}

Next, we study the influence of the ISRF and CRIR on the properties of the filaments (right panel of Fig.~\ref{fig:species}). Increasing the CRIR naturally results in a higher ionisation fraction of gas. This is reflected by the abundances of H$^+$ and C$^+$, which are 1 - 2 orders of magnitude higher in runs with \mbox{CRIR = $1 \times 10^{-16}$ s$^{-1}$}. Also $T_\text{gas}$ shows a slight increase with increasing CRIR, which is due to the larger amount of energy released by various dissociation reactions caused by cosmic rays. Increasing the strength of the ISRF only marginally affects the chemical composition of the gas. However, $T_\text{gas}$ and $T_\text{dust}$ are increased by a few Kelvin, which is most likely due to the increased photoelectric heating. Interestingly, we find that for all runs $T_\text{dust}$ decreases towards the centre of the filaments. We attribute this to the progressive attenuation of the ISRF -- mainly responsible for dust heating -- towards the centre of the filament \citep[see also][]{Clark13}. 

\section{Discussion and conclusions}
\label{sec:dis}

\subsection{Physical interpretation}

Assuming a polytropic relation $T_\text{dust} \propto \rho^{\gamma - 1}$, we find $\gamma$ to be in a range from 0.9 to 0.95, which is in good agreement with HERSCHEL observations by \citet{Palmeirim13}. We emphasize that $T_\text{gas}$ and $T_\text{dust}$ are markedly different, which requires independent measurements for both quantities in observations. We attribute this to the fact that the collisional coupling between gas and dust at \mbox{$n \sim 10^5 - 10^6$ cm$^{-3}$} is not yet strong enough to assure similar temperatures.

It can also be seen that the ratio $R = \frac{n_{\text{CO}}}{n_\text{H$_2$}}$ is not constant along the radial direction. There is a strong decline of $R$ with increasing radius of about 2 orders of magnitude, which is due to the fact that the formation of CO happens at somewhat smaller radii (i.e. higher gas column densities) than that of H$_2$. Our work therefore suggests that caution is recommended when using CO line intensities (i.e. the X-factor) to obtain total gas masses. In the centre of the filament, however, $R$ is comparable for all our runs with a value around $1.5 \times 10^{-4}$, and thus in good agreement with observations.

\subsection{Numerical cost/Performance}

The simulations were carried out on computing nodes with 2 - 5 GB memory per CPU, which corresponds to state-of-the-art computing nodes at modern supercomputing facilities. We compare our runs with an isothermal ($T$ = 15 K) simulation without a chemical network. We find that the inclusion of the network increases the computation costs per timestep by roughly a factor of 5 -- 7.

\subsection{Conclusions and outlook}

We present first results of simulations of star forming filaments using -- to our knowledge -- the largest chemical network ever applied in 3D-MHD simulations \citep[see also][]{Seifried15b}, combining the versatile chemistry package KROME \citep{Grassi14} with the TreeCol algorithm \citep{Clark12,Wunsch15}. We show that in terms of memory consumption such simulations are feasible on modern supercomputers. The results appear to be promising and in good agreement with observational results.

This work paves the way for many future applications: The detailed chemical modelling will allow us to produce synthetic observations of several molecular lines as well as of continuum emission. Furthermore, the simulations can improve our understanding of the X-factor required for the mass determination of gaseous objects. Moreover, the network will allow us to study in detail the evolution of other molecules like H$_2$O or the cosmic ray tracer H$_3^+$.

%%-----------------------------
%%      your bibliography
%%-----------------------------

\end{document}